\documentclass[twocolumn, twocolappendix]{aastex701}
\usepackage{graphicx}
\usepackage{bm}
\usepackage{hyperref}
\usepackage{float}
\usepackage{amsmath, amssymb}
\usepackage{xcolor}
\usepackage{enumitem}

\newcommand{\Qg}{Q_{\mathrm{g}}}
\newcommand{\Qs}{Q_{\mathrm{s}}}

\newcommand{\hg}{h_{\mathrm{g}}}
\newcommand{\hs}{h_{\mathrm{s}}}
\newcommand{\Sg}{\Sigma_{\mathrm{g}}}
\newcommand{\Sgo}{\Sigma_{\mathrm{g}0}}
\newcommand{\Ss}{\Sigma_{\mathrm{s}}}
\newcommand{\Sso}{\Sigma_{\mathrm{s}0}}
\newcommand{\Pg}{P_{\mathrm{g}}}
\newcommand{\Phitot}{\Phi_{\mathrm{tot}}}
\newcommand{\cg}{c}
\newcommand{\yc}{y_{\mathrm{c}}}
\newcommand{\dhSg}{\delta \hat{\Sigma}_{\mathrm{g}}}
\newcommand{\hu}{\hat{u}}
\newcommand{\hv}{\hat{v}}
\newcommand{\dhf}{\delta \hat{f}}
\newcommand{\dhSs}{\delta \hat{\Sigma}_{\mathrm{s}}}
\newcommand{\kzero}{k_0}
\newcommand{\Kpot}{K_{\mathrm{pot}}}

\usepackage[normalem]{ulem}

\allowdisplaybreaks[4]

\makeatletter
\renewcommand*\env@matrix[1][\arraystretch]{%
  \edef\arraystretch{#1}%
  \hskip -\arraycolsep
  \let\@ifnextchar\new@ifnextchar
  \array{*\c@MaxMatrixCols c}}
\makeatother

\begin{document}

\title{Swing amplification in star-gas disks}

\author[0000-0002-4099-4359]{Abhishek Hegade K. R.}
\affiliation{Princeton Gravity Initiative, Princeton University, Princeton, NJ 08544, USA}
\email{ah4278@princeton.edu}

\author[0000-0002-5861-5687]{Chris Hamilton}
\affiliation{Department of Astrophysical Sciences, Princeton University,
4 Ivy Lane, Princeton, NJ 08544, USA}
\affiliation{Institute for Advanced Study,
Einstein Drive, Princeton, NJ 08540, USA}
\email{chamilton@princeton.edu}

\begin{abstract}
Recent JWST and ALMA observations have revealed stellar bars and spirals in gas-rich galactic disks at redshifts as high as $z \simeq 4$.  The simplest theoretical paradigm we have for understanding such non-axisymmetric features is the linear theory of swing amplification (SA) in the 2D shearing sheet.  However, while the SA mechanism in a gaseous shearing sheet was first studied in 1965, and that in a collisionless stellar sheet in 1966, the coupled star-gas linear SA equations have never been solved explicitly. Here we write down these equations and use them to study the evolution of non-axisymmetric swinging waves in stable disks. 
We find that  waves are often amplified temporarily by factors of $10-100$ or more, and that the maximum \textit{non-axisymmetric} amplification factor is tightly correlated with the system's distance from the \textit{axisymmetric} stability boundary in the $(1/Q_\mathrm{s}, 1/Q_\mathrm{g})$ plane.
The true star-gas behavior differs significantly from the `two-fluid' idealizations used in the past, because phase mixing of the collisionless stellar component acts as a sink of perturbation energy. 
Closely analogous results hold for finite-thickness disks except for the shifting of the stability boundary. 
We provide a \texttt{python} code that calculates the SA factors and maximally-amplified wavelength given the background disk parameters.
\end{abstract}


\section{Introduction}
\label{sec:introduction}
Observations utilizing the James Webb Space Telescope (JWST) and the Atacama Large Millimeter/submillimeter Array (ALMA) have revealed pronounced non-axisymmetric features, such as stellar bars and spiral arms, in galactic disks at redshifts as high as $z \simeq 4$ \citep{guo2023first,kuhn2024jwst,boogaard2026stellar}.  These observations promise to be of great importance for the study of galaxy evolution --- they might tell us when disks settled into the rotation-dominated structures we see today, how the Hubble sequence emerged, and so on \citep{smethurst2025galaxy}. However, extracting physical information from the observations is difficult.

One major issue is that, at these redshifts, galaxies were sufficiently gas rich \citep{tac20} that there was significant coupling between the gaseous and stellar disks, but we do not yet have a deep understanding of coupled star-gas dynamics. 
On the analytic side, most canonical works on stellar dynamics (e.g., \citealt{BT}) treat star-gas coupling rather crudely or simply ignore it altogether.
On the numerical side, high-resolution global simulations of isolated, near-equilibrium galaxies with large gas fractions are only beginning to reach maturity (e.g, \citealt{bland2023rapid,blandhawthorn2024}).   Because of these limitations, we possess no quantitative theory of the formation and properties of galactic bars and/or spiral arms in gas-rich  environments with which to interpret the data. 

From where might such a theory arise? The simplest paradigm we have for understanding the generation of collectively-excited non-axisymmetric features in galactic disks is the linear theory of swing amplification (SA) in a local 2D patch of the disk, a.k.a. the `shearing sheet'.
Historically, the mechanics of SA in the shearing sheet were developed independently for single-component systems: \citet{GLB1965} explored a purely gaseous shearing sheet, while \citet{JT66} explored its collisionless stellar analogue. 
\citet{jog1992swing} went a step further, and studied SA in a `star-gas' sheet, but her `stars' were actually treated as a fluid whose sound speed mimicked the velocity dispersion of a stellar population. Later, \citet{rafikov2001local} derived the local axisymmetric instability criterion for a multi-component disk, correctly treating the stars as collisionless rather than fluid, but he did not study non-axisymmetric perturbations. Thereafter \cite{Kim2007} investigated fragmentation and turbulence in realistically thick shearing patches of star-gas disks. They began analytically by extending Rafikov's axisymmetric dispersion relation to disks of nonzero thickness, but studied the non-axisymmetric part of the problem via numerical simulations (for the current state of the art see, e.g., \citealt{kim2023introducing}). It remains the case that the linear theory of non-axisymmetric SA in star-gas disks --- with the stars treated as collisionless --- has never been studied explicitly.

In this paper we fill precisely this theoretical gap, studying the amplification of non-axisymmetric swinging waves in 2D razor-thin star-gas disks, treating the stars as collisionless. We also study disks of finite thickness following the effective prescription from \cite{Kim2007}.
The major caveats to our work are that we focus on \textit{linear} theory in \textit{local} patches of \textit{two-component} disks with no \textit{turbulence}, no \textit{magnetic fields}, and no \textit{complicated thermodynamics}.
Each of these assumptions can (and should) be relaxed in the future as we work towards a realistic theory of coupled star-gas disk dynamics.

The structure of the paper is as follows. 
In \S\ref{sec:theory}, we define the background state and write down the linear perturbation master equations for the combined star-gas system.
In \S\ref{sec:single_wave}, we use these equations to study the evolution of impulsively-excited non-axisymmetric swinging waves. 
We discuss the implications and limitations of our results and conclude in \S\ref{sec:discussion}.
We keep the mathematical machinery to a minimum in the main text, but many details are gathered in the Appendices. 
All numerical results shown in this paper can be reproduced using our \texttt{python} code \citep{github_code}. 

\section{Linear theory}
\label{sec:theory}
In this section we derive the linear equations for a coupled star-gas shearing sheet in 2D (\S\ref{sec:2D}),  
write down a simple finite-thickness generalization thereof (\S\ref{sec:3D}), discuss the corresponding axisymmetric stability criteria (\S\ref{sec:axisymmetric}), and perform a non-trivial code test by calculating the response to a corotating massive cloud (\S\ref{sec:cloud}).

\subsection{Linear theory in the 2D shearing sheet}
\label{sec:2D}
We begin by considering a 2D disk confined to the $z=0$ plane of the usual polar coordinate system $(\varphi, R, z)$, with $R=0$ being the disk center. 
The total gravitational potential of the system is the sum of gravitational potentials generated by the gas, the stars, and any external potentials:
\begin{align}
   \Phi_{\mathrm{tot}} = \Phi_\mathrm{g}+\Phi_\mathrm{s} + \Phi_{\mathrm{ext}}.
\end{align}
We will always take our disk to be initially unperturbed, in which case each of these contributions to the potential is axisymmetric and time-independent. The total potential is then some fixed 
$\Phi_\mathrm{tot,0}$, and from this we can derive the background rotation curve, azimuthal frequency profile $\Omega(R)$, etc. in the usual way.  

We focus on a local shearing sheet centered on a point with radius $R=\overline{R}$, rotating at the local circular frequency $\Omega(\overline{R})$, which we simply refer to as $\Omega$ hereafter.
The Oort constants $A$, $B$ and the local epicyclic frequency $\kappa$ have the usual definitions
\begin{align}
    A \equiv -\frac{1}{2} \frac{d \Omega}{d \ln R} \bigg\vert_{R = \overline{R}},
    \quad \quad  B \equiv A - \Omega = -\frac{\kappa^2}{4\Omega}.
\end{align}
Within the sheet we introduce local coordinates
\begin{align}
    x \equiv R - \overline{R} \,,\quad \quad 
    y \equiv \overline{R} \left( \varphi - \Omega t \right).
\end{align}
and the corresponding velocities $v^x$, $v^y$.

In Appendix \ref{sec:appendix_derivation} we consider the equations describing the coupled gas-star dynamics 
in these coordinates for small $\vert x \vert, \, \vert y\vert  \ll \overline{R}$.
The gas satisfies momentum and continuity equations~\citep{Ryu-Goodman-1992} and we assume no viscosity. The stellar distribution function $f$ satisfies the collisionless Boltzmann equation \citep{Binney_2020}. The two components are coupled through the gravitational potential $\Phi_\mathrm{tot}$. The contribution of each component to this potential satisfies the Poisson
equation
\begin{align}\label{eq:Poisson-eqn}
    \nabla^2 \Phi_i = 4\pi G \Sigma_{i} \delta(z), 
\end{align}
where $\nabla^2$ is the 3D Laplacian operator, $i=\{ \mathrm{s,g}\}$, and the surface densities in gas and stars are $\Sigma_{\mathrm{g}}$ and
\begin{align}
    \Sigma_{\mathrm{s}} \equiv\int_{-\infty}^{\infty}dv_x   \int_{-\infty}^{\infty} \, dv_y \, f,
\end{align}
respectively. 

We then linearize these equations around an unperturbed background state  (denoted by subscript $0$) with constant (i.e., both time-independent and homogeneous) stellar and gaseous surface densities, gas pressure, and gravitational potential $\Phi_\mathrm{tot,0}$; for the background gas velocity profile we take
\begin{align}
    v_0^x = 0\,,\quad \quad v_0^y = -2 A x \,,
\end{align}
and for the stellar distribution function we use
\begin{align}\label{eq:Vlasov-equilibrium}
    f_0 = \frac{\Sigma_{\mathrm{s}0}}{2\pi \sigma \sigma_y} e^{-{(v^{x})^2}/({2\sigma^2} )}
e^{   -
    (v^y-v^y_{0})^2/({2 \sigma_y^2})}
    \,,
\end{align}
where $\sigma$ is the radial velocity dispersion and $\sigma_y \equiv \sigma {\kappa}/{(2 \Omega)}$ \citep{JT66,Binney_2020}.
We decompose all quantities into this background state plus small perturbations (denoted by $\delta$), and discard all nonlinear terms. We assume gas pressure and surface density perturbations are related by
\begin{align}
    \delta \Pg = \cg^2 \delta \Sg \,,
    \label{eqn:isothermal}
\end{align}
where $\cg$ is the constant speed of sound. 

Having done this we transform to a new shearing coordinate system $(x_\mathrm{c},y_\mathrm{c})$ that co-moves with the unperturbed fluid flow:
\begin{align}\label{eq:comoving-frame-transform}
    x_\mathrm{c} \equiv x, \quad \quad \yc \equiv y + 2 A x t.
\end{align}
(Note that, without loss of generality, we have singled out $t=0$ as a special time when the two coordinate systems are exactly aligned with each other.) We can Fourier transform the fluctuations in the shearing coordinates according to
\begin{align}
\label{eq:fourier-transformation-convention}
&\delta \hat{X}(k_{x_\mathrm{c}}, k_{y_\mathrm{c}}, t) \equiv 
\int \frac{d x_\mathrm{c}}{2\pi}\frac{d y_\mathrm{c}}{2\pi} \delta X(x_\mathrm{c},y_\mathrm{c}, t) 
\,e^{-i (k_{x_\mathrm{c}} x_\mathrm{c} + k_{y_\mathrm{c}} y_\mathrm{c})},
\end{align}
and in linear theory the individual Fourier modes $\delta \hat{X}$ evolve independently \citep{GLB1965,JT66}. In the non-shearing $(x,y)$ coordinate system,
these independent modes look like shearing waves with wavenumber
\begin{equation}
    k_x = k_{x_\mathrm{c}}+2Ak_{y_\mathrm{c}} t, \quad \quad k_y = k_{y_\mathrm{c}}.
    \label{eqn:shearing_waves}
    \end{equation}
Finally, from Eq. \eqref{eq:Poisson-eqn} the density and potential Fourier modes are related via
\begin{align}
    \delta \hat{\Phi}_{i} \equiv 
     -
    \frac{2 \pi G \delta \hat{\Sigma}_{i} }{k(t)}   
    \,,
    \label{eqn:poisson-solution}
\end{align}
where
\begin{align}
    &k(t) \equiv \sqrt{k_{x}^2+k_{y}^2} = \sqrt{( k_{x_{\mathrm{c}}} + 2 A k_{y_{\mathrm{c}}} t)^2 + k_{y_{\mathrm{c}}}^2}.
    \label{eqn:koft}
\end{align}

Using these identities the governing equations reduce to two coupled equations for 
gaseous and stellar potential perturbation Fourier components (see Appendix \ref{sec:appendix_derivation}):
    \begin{align}\label{eq:gas-master-eqn}
    &\partial_t^2 \delta \hat{\Phi}_{\mathrm{g}} 
    +
    S^2(t)
    \delta \hat{\Phi}_{\mathrm{g}}
    =
    2\pi G \Sigma_{\mathrm{g}0} k(t)
    \left[ 
    \delta \hat{\Phi}_{\mathrm{s}}
    +
    \delta \hat{\Phi}_{\mathrm{ext}}
    \right]
    \,,\\
    \label{eq:stars-master-eqn}
    & \delta \hat{\Phi}_{\mathrm{s}}(t)
    = 
    \int_{t_\mathrm{i}}^t dt' \Kpot(t,t')  
    \nonumber\\
    &\quad \quad \quad \quad \quad \quad  \times 
    \left[ \delta \hat{\Phi}_{\mathrm{g}}(t') +  \delta \hat{\Phi}_{\mathrm{s}} (t') + \delta \hat{\Phi}_{\mathrm{ext}} (t') \right],
\end{align}
where the time-dependent `spring constant' $S(t)$ is given in Eq. \eqref{eqn:spring_constant}, $t_\mathrm{i}$ is some reference time when the disk was unperturbed,
and the kernel $\Kpot(t,t')$ is given in Eq. \eqref{eqn:potential_kernel}.  
It is easy to check that in the limit of a purely gaseous disk $(\Sigma_{\mathrm{s}0}=0)$ we recover the same equations as \citet{GLB1965}, while for a purely stellar disk $(\Sigma_{\mathrm{g}0}=0)$ we recover those of \cite{JT66}.


\subsection{Finite-thickness disks}
\label{sec:3D}

The discussion so far holds only for razor-thin (2D) disks. However, real disks have non-zero thickness and this can change the linear response properties significantly \citep{Kim2007,meidt2022molecular}. 

A fully self-consistent 3D analysis of star-gas swing amplification is beyond the scope of this paper. Instead, following \cite{Kim2007}, we adopt a simple effective prescription in which the horizontal dynamics of both components is governed by the perturbed gravitational potential evaluated at the disk midplane ($z=0$). We assume that each component has an exponential vertical density profile with scale height $H_i$, where $i\in\{\mathrm{g},\mathrm{s}\}$, and that the perturbations have the same fixed vertical shape as the corresponding background component. Then a surface density perturbation $\delta\hat{\Sigma}_i$ corresponds to a 3D density perturbation
\begin{align}
    \delta\hat{\rho}_i(z)
    =
    \frac{\delta\hat{\Sigma}_i}{2H_i}
    e^{-|z|/H_i}.
\end{align}
Solving the Poisson equation
for this source and evaluating the result at $z=0$ gives the \textit{midplane} potential
\begin{align}
    \delta\hat{\Phi}_i
    =
    -
    \frac{2\pi G\,\delta\hat{\Sigma}_i}
    {k(t)\left[1+H_i k(t)\right]}.
    \label{eqn:thick-midplane-poisson}
\end{align}
Thus the finite thickness enters as a reduction factor, $(1+H_i k)^{-1}$, according to the scale height of the component that sources it. 

Once we replace Eq.~\eqref{eqn:poisson-solution} with Eq.~\eqref{eqn:thick-midplane-poisson}, the rest of the derivation from \S\ref{sec:2D} carries through essentially unchanged. The resulting effective thick-disk master equations are given in Appendix~\ref{appendix:thick-disk}.

\subsection{Axisymmetric limit}
\label{sec:axisymmetric}

\begin{figure}
    \centering
    \includegraphics[width=0.99\linewidth]{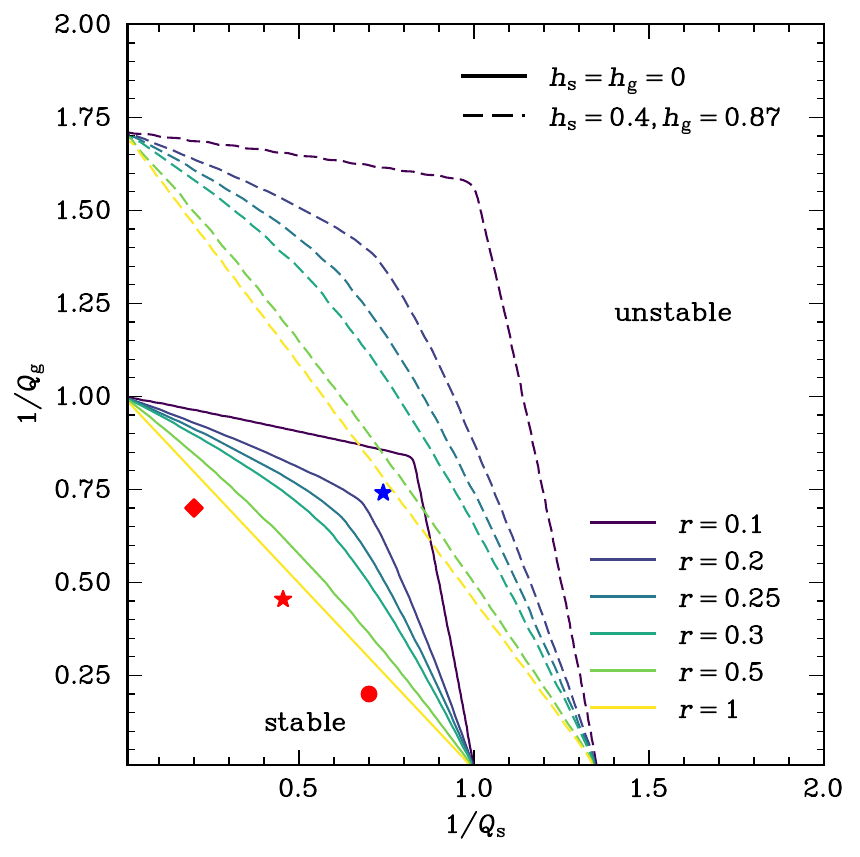}
    \caption{Stability of star-gas disks to local axisymmetric perturbations. Given a value of $r=c/\sigma$ (denoted by colors), regions below the curves are stable and those above are unstable according to the criterion \eqref{eq:rafikov-stability}.
    Solid lines are for razor-thin disks and dashed lines are for thicker disks with $h_\mathrm{s}=0.4$ and $h_\mathrm{g}=0.87$. The various red and blue symbols denote specific locations in the $(1/Q_\mathrm{s}, 1/Q_\mathrm{g})$ plane that we will study later.}
    \label{fig:rafikov_dispersion}
\end{figure}

We can take the axisymmetric limit of the linearized equations \eqref{eq:gas-master-eqn}-\eqref{eq:stars-master-eqn} by setting $k_y=0$.  We can then derive an axisymmetric stability criterion by looking for solutions with time dependence $ \propto e^{-i\omega t}$ with Im $\omega > 0$. This is a standard exercise so we will not repeat the details here, but just write down the key result. 

We introduce the dimensionless Toomre $Q$ parameters\footnote{Note that our definition of $\Qs$ agrees with that of~\cite{Kim2007}, but differs from that of \cite{rafikov2001local} by a factor of $\pi/3.36$. }
\begin{align}
    &\Qs \equiv \frac{\kappa \sigma}{3.36 G \Sigma_{\mathrm{s}0}} \,,\quad\quad \Qg \equiv \frac{\kappa c}{\pi G \Sigma_{\mathrm{g}0}},
\end{align}
the dimensionless ratio $r$
\begin{equation}
    r = \frac{c}{\sigma} = \frac{3.36}{\pi}\frac{Q_\mathrm{s}}{Q_\mathrm{g}}\frac{\Sigma_{\mathrm{g}0}}{\Sigma_{\mathrm{s}0}},
\end{equation}
the dimensionless disk thicknesses
\begin{align}
    &h_\mathrm{g} \equiv \frac{\kappa H_\mathrm{g}}{c}, \quad\quad 
    h_\mathrm{s} \equiv \frac{\kappa H_\mathrm{s}}{\sigma},
\end{align}
and the wavenumber
\begin{equation}
k_{\sigma} \equiv \frac{\kappa}{\sigma}.
\label{eqn:kcrit}
\end{equation}
Given the five dimensionless numbers  $(\Qs,Q_\mathrm{g}, r, \hs, \hg )$ and the wavenumber $k_{\sigma}$, a star-gas disk is unstable to local axisymmetric perturbations with wavenumber $k_x$ if~\citep{rafikov2001local,Kim2007}:
\begin{align}\label{eq:rafikov-stability}
    &\frac{2}{Q_\mathrm{s}}
    \frac{\pi}{3.36 u(1+u \hs)}
    \left[1 - e^{-u^2} I_0 \left(u^2 \right) \right]
    \nonumber\\
    &\quad \quad 
    \quad \quad +
    \frac{2}{Q_\mathrm{g}}
    \frac{r u}{(1+ u^2 r^2)(1 + \hg  u r)}
    >
    1\,,
\end{align}
where $u=\vert k_x\vert /k_{\sigma}$ and $I_0$ is a modified Bessel function of the first kind.

In Fig.~\ref{fig:rafikov_dispersion} we plot the stability criterion \eqref{eq:rafikov-stability} in the $(1/\Qs, 1/\Qg)$ plane.
The solid lines denote the stability boundary for razor-thin disks ($h_\mathrm{g}=h_\mathrm{s}=0$) with different $r$ (denoted by colors), while dashed lines show the same results for thick disks with $\hs = 0.4$ and $\hg=0.87$. Given a value of $r$, regions below (above) the curve correspond to disks that are locally stable (unstable) to axisymmetric perturbations. The various red and blue symbols in this panel correspond to specific locations in the $(1/Q_\mathrm{s}, 1/Q_\mathrm{g})$ plane that we will study numerically.


\subsection{Numerical implementation and an example}
\label{sec:cloud}

In general we must solve equations \eqref{eq:gas-master-eqn}-\eqref{eq:stars-master-eqn} numerically. We do this in practice by first recasting them as a Volterra integral equation and then integrating forward in time using the trapezoidal rule.  We give details of our numerical scheme in Appendix~\ref{appendix:num-scheme}.

\begin{figure}
    \centering
    \includegraphics[width=0.95\linewidth]{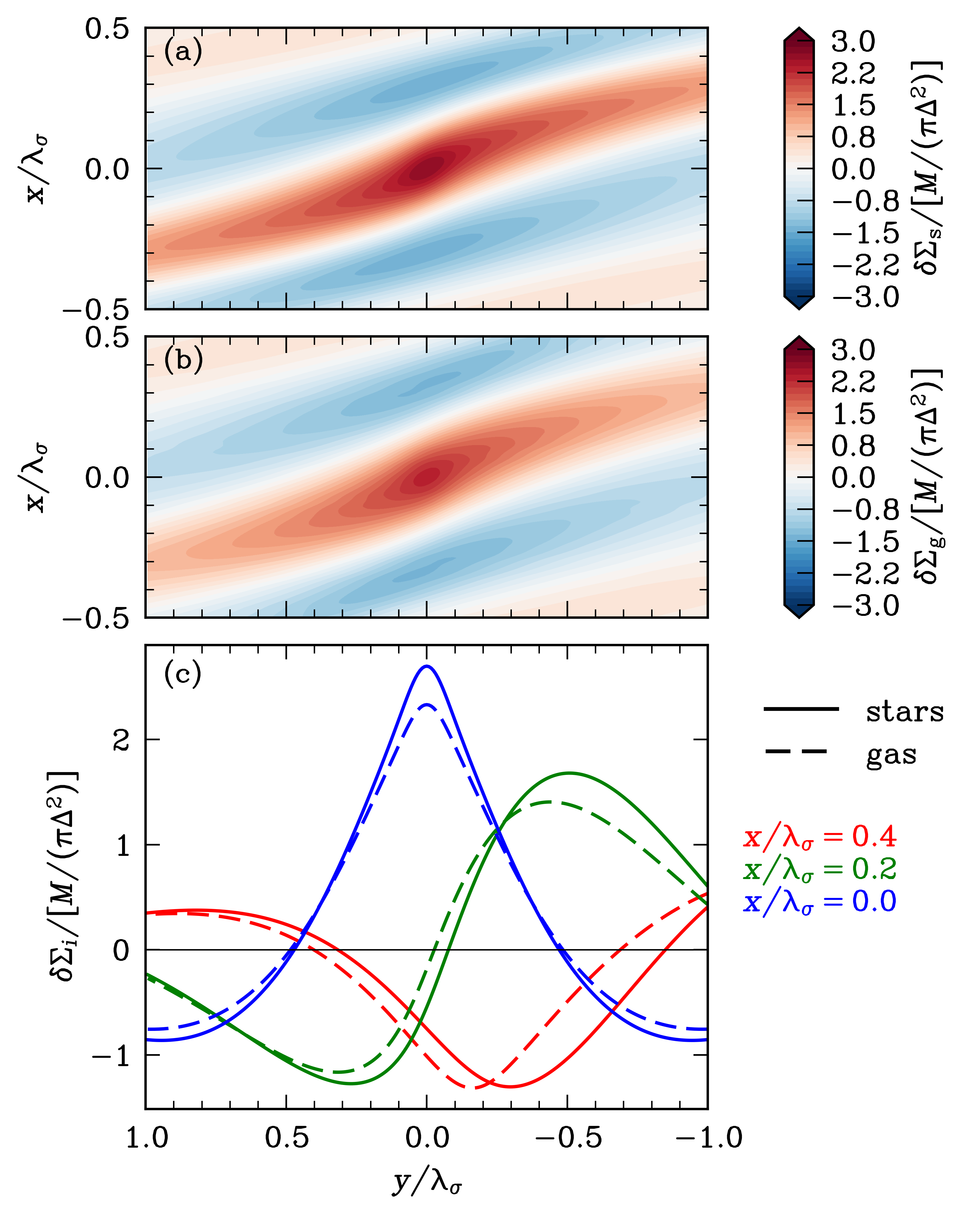}
    \caption{
    Time-asymptotic linear response of a razor-thin shearing sheet to a corotating massive cloud at its origin, for a disk with parameters
$Q_{\rm s} = Q_{\rm g}=2.2$ (corresponding to the location of the red star in Fig.~\ref{fig:rafikov_dispersion}), and $r=1.0$.
    Panels (a) and (b) show the stellar and gaseous surface density perturbations respectively in units of
    $M/(\pi\Delta^2)$, where $M$ is the cloud mass and $\Delta=0.05\,\lambda_{\sigma}$ with $\lambda_{\sigma} = 2\pi/k_{\sigma}$ (Eq. \eqref{eqn:kcrit}).
Panel (c) shows some slices of these surface density perturbations as a function of $y$ at fixed $x$; solid lines are for stars and dashed lines for gas.
    }
    \label{fig:cloud}
\end{figure}

As a check of our code we performed a classic 2D
calculation \citep{JT66}: the time-asymptotic linear response of a shearing sheet to a
corotating massive cloud. We model the cloud via the
external surface density
\begin{align}
    \Sigma_{\mathrm{ext}}(x,y)
    =
    \frac{M}{2\pi\Delta^2} e^{-({x^2+y^2})/({2\Delta^2})},
\end{align}
where $M$ is the cloud mass $\Delta$ is its characteristic radius; the details of the ensuing calculation are given in
Appendix~\ref{appendix:cloud_details}.  We first made sure that we could reproduce Fig.~9 of \citet{Binney_2020}, which is for a star-only disk. Then we turned to star-gas disks. In Fig.~\ref{fig:cloud} we show the result of a calculation with 
$Q_{\rm s} = Q_{\rm g}=2.2$, $r=1.0$ and $\Delta = 0.05\lambda_{\sigma}$, where 
\begin{equation}
    \lambda_{\sigma} \equiv \frac{2\pi}{k_\sigma} = \sqrt{2}\pi a,
\end{equation}
where $a$ is the rms epicyclic amplitude of the stars. This (stable) model is marked by the red star in Fig.~\ref{fig:rafikov_dispersion}. 
Panels (a) and (b) of Fig.~\ref{fig:cloud} show contours of the resulting density perturbations in stars and gas respectively while panel (c) shows slices through these density fields at fixed $x$.  In each case the units are $M/(\pi\Delta^2)$.
We see from this figure that the self-gravity of the system can create a steady wake around the cloud that is significantly more massive and spatially extended than the cloud itself.  Stellar and gaseous wakes are of similar shapes, and tend to reinforce one another gravitationally.
Since we are dealing with an idealized linear theory there are no shocks or dissipation in our gas so we do not produce the very sharp gas features as in, e.g., Fig. 9 of \cite{Kim2007}.


\section{Amplification of a  swinging wave}
\label{sec:single_wave}

In this section we study the temporal evolution of swinging waves.  In each case we excite the disk impulsively at time $t=t_\mathrm{i}$, using an external density perturbation that is razor-thin with surface density Fourier component $\delta \hat{\Sigma}_\mathrm{imp}(k_{x_\mathrm{c}},k_{y_\mathrm{c}})$ in the shearing coordinates (see Eq.~\eqref{eq:fourier-transformation-convention}). 
We know that, when viewed in the non-shearing coordinates $(x,y)$,
each of these Fourier components will correspond to a swinging wave with azimuthal wavenumber $k_y = k_{y_\mathrm{c}}$ and radial wavenumber $k_{x} = k_{x_\mathrm{c}}+2Ak_{y_\mathrm{c}} t$ (Eq.~\eqref{eqn:shearing_waves}).
Following \cite{Binney_2020}, and without loss of generality, we choose the alignment of our shearing coordinate system such that $k_{x_c}=0$. This guarantees that, when viewed in the unsheared coordinates $(x,y)$, the wave crests are leading for $t<0$, aligned radially at $t=0$, and trailing for $t>0$. The wave's initial radial structure in the unsheared coordinates is determined by the choice of perturbation time, $k_x(t_\mathrm{i})=2Ak_{y_\mathrm{c}}t_\mathrm{i}$.
In practice, only a narrow range of values $-2.5 \lesssim \kappa t_\mathrm{i} / \pi \lesssim -0.5$ can give rise to a vigorous self-gravitating response (see, e.g., Fig. 3 of \citealt{Binney_2020}).
\subsection{A single azimuthal wavenumber}
\label{sec:single}

\begin{figure*}
    \centering
    \includegraphics[width=0.95\linewidth]{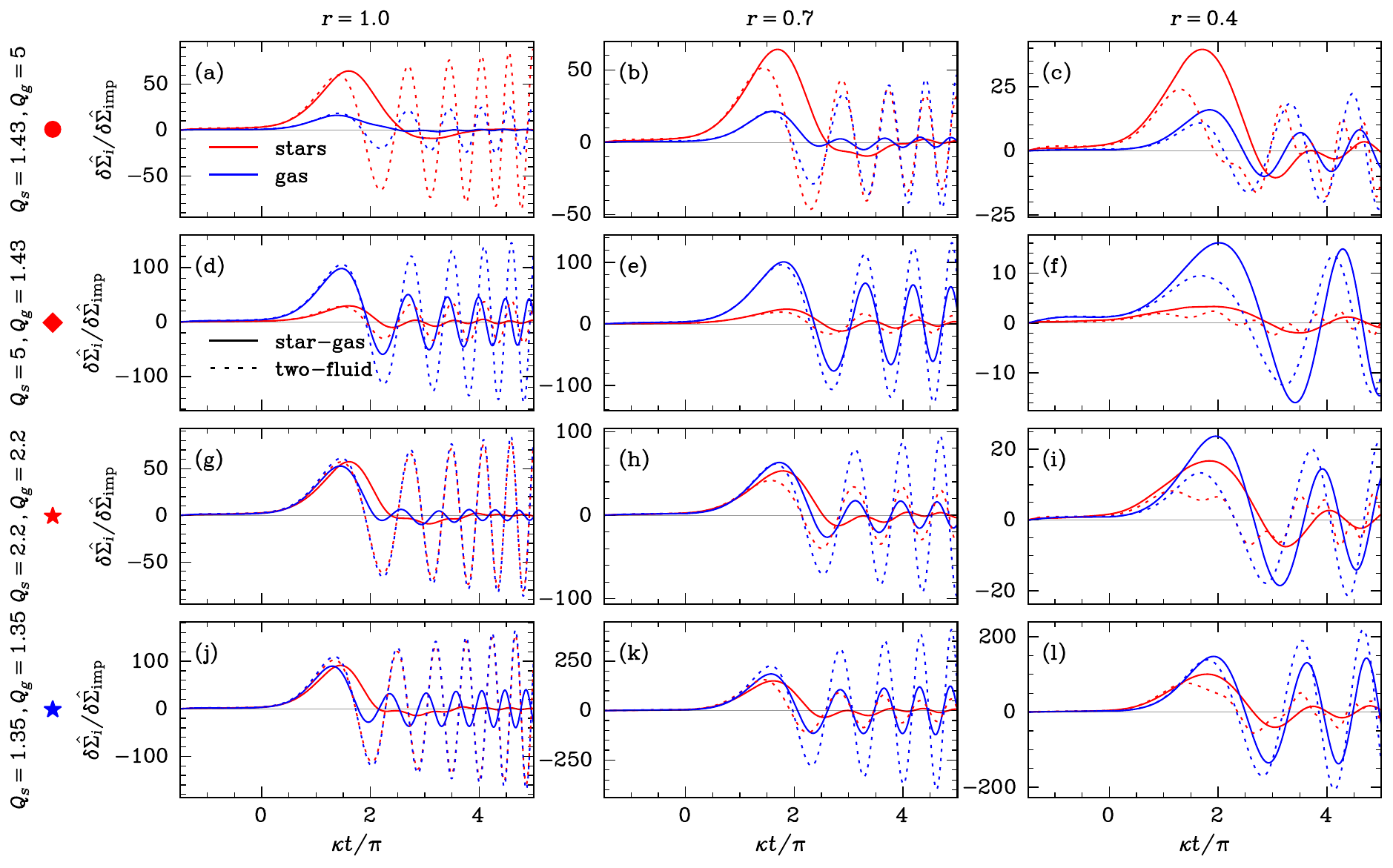}
    \caption{Amplification of swinging waves induced by an impulsive surface density perturbation $\delta\hat{\Sigma}_{\mathrm{imp}}$ with $k_{y_\mathrm{c}}/k_{\sigma} = 0.5$ at $\kappa t_\mathrm{i}/\pi = -1.5$. Stellar (gaseous) responses are shown with solid red (blue) lines. Each row is for a fixed pair of $\Qs, \Qg$ values (corresponding to the symbols in Fig.~\ref{fig:rafikov_dispersion}); the first three rows are for razor-thin disks ($h_\mathrm{s}=h_\mathrm{g}=0$) while the fourth row is for a thicker disk ($h_\mathrm{s}=0.4, \,\,h_\mathrm{g}=0.87$). Each column is for a different ratio $r=c/\sigma$. The dashed lines show the results of an effective two-fluid calculation in which the collisionless stellar component is replaced by a fluid with sound speed $\sigma$.
    }
    \label{fig:single_k_swing_amplification}
\end{figure*}

First we illustrate the response to a perturbation with fixed $k_{y_{\mathrm{c}}}/k_{\sigma} =0.5$ and $\kappa t_\mathrm{i}/\pi =-1.5$. 
In Fig.~\ref{fig:single_k_swing_amplification}, with solid lines we show the resulting stellar (red) and gaseous (blue) surface density responses, $\delta\hat{\Sigma}/\delta\hat{\Sigma}_{\mathrm{imp}}$, to this perturbation for twelve different disks.   As indicated in the figure, these twelve disks are generated from three different dimensionless ratios $r$ (different columns), and four pairs of Toomre parameters $\Qs$ and $\Qg$ (different rows) each of which corresponds to one of the symbols shown in the stability diagram in Fig.~\ref{fig:rafikov_dispersion}. The top three rows show results of 2D calculations, while the bottom row is for the finite-thickness calculation\footnote{To be explicit, in the finite-thickness calculation we assume that the external perturbation is exactly the same as in the 2D case, i.e., it is due to some razor-thin externally-applied surface density.} described in \S\ref{sec:3D} with $h_\mathrm{s} = 0.4$, $h_\mathrm{g}=0.87$.  Each of these disks is stable according to the axisymmetric criterion \eqref{eq:rafikov-stability}. In addition, in each panel we show with dashed lines the results of an analogous two-fluid calculation \citep{jog1992swing} in which we treat the `stars' as a fluid with sound speed $\sigma$.

First we focus on the solid lines in the top row of Fig.~\ref{fig:single_k_swing_amplification}, for which $Q_\mathrm{s} = 1.43$ and $Q_\mathrm{g}=5$ (corresponding to the red circle in Fig.~\ref{fig:rafikov_dispersion}). In this limit the stars are strongly self-gravitating while the gas is not. The density evolution consists of a single major amplification event within the first epicyclic period after the wave becomes trailing ($0 < \kappa t/\pi < 2$), followed by damped oscillations. The 
maximum amplification factor in the stars is $\sim 50$ in all three panels. Moreover, although the gas can hardly self-gravitate for this value of $Q_\mathrm{g}$, the perturbed potential of the stars is enough to amplify the gaseous perturbation by a factor $\sim 20$. 

This story is basically reversed in the second row of Fig.~\ref{fig:single_k_swing_amplification}, for which $Q_\mathrm{s} = 5$ and $Q_\mathrm{g}=1.43$ (red diamond in Fig.~\ref{fig:rafikov_dispersion}). Again the cool component (gas) is amplified significantly (this time by up to a factor $\sim 100$) while the hotter component (stars) get taken along for the ride, still being amplified but by a factor a few times smaller than the cool component.

The third row in Fig.~\ref{fig:single_k_swing_amplification} shows what happens when stars and gas truly co-operate. In this case $Q_\mathrm{s}=Q_\mathrm{g}=2.2$ (red star in Fig.~\ref{fig:rafikov_dispersion}), so neither component ought to be strongly self-gravitating on its own (see, e.g., Fig. 4 of \citealt{Binney_2020}). Yet we see from panels (g)-(i) of Fig.~\ref{fig:single_k_swing_amplification} that the combined response of the two components is nearly in-phase, and that this leads to amplification by as much as a factor $\sim 50$. 
A closely analogous story holds in the fourth row of Fig.~\ref{fig:single_k_swing_amplification} which is for a disk of finite thickness with $Q_\mathrm{g} = Q_\mathrm{s} = 1.35$ (blue star in Fig.~\ref{fig:rafikov_dispersion}). The reduction in Toomre $Q$ values is necessary to get any significant amplification at all --- otherwise the system is far from the stability boundary --- but qualitatively the dynamics is the same.

Next, it is interesting to compare these results with those of the two-fluid prescription utilized by \cite{jog1992swing}, which we show with dashed lines in Fig.~\ref{fig:single_k_swing_amplification}. In each panel we see that the two-fluid model does a reasonable job of reproducing the early stages of the response for both gas and stars.   However, after the first peak has been reached, we begin to find significant discrepancies between two-fluid and proper star-gas calculations. For instance, the two-fluid approach predicts very long-lived oscillations in both gas and stars.\footnote{Some of these oscillations are still growing at the latest time shown here $\kappa t/\pi = 5$, but we have checked that each two-fluid disk is in fact stable, and oscillates neutrally over much longer timescales.} This is because in the absence of any viscosity, a two-fluid disk has no way to lose energy, so non-axisymmetric oscillations can in principle live forever. On the other hand, a true collisionless stellar disk exhibits \textit{phase mixing} --- perturbations to its distribution function wind to ever smaller scales in velocity space and the resulting density perturbations (which are integrals over velocity) decay with time. 
Thus the stars are a `sink' of perturbation energy\footnote{Technically, phase mixing in a perfectly collisionless system is reversible nonlinearly \citep{chiba2025galactic}, but a tiny amount of dissipation at small phase-space scales (or effective dissipation through coarse-graining thereof) renders it irreversible, which we interpret as the system `heating up' \citep{tremaine1986h,dehnen2005phase}.} and since gas and stars are coupled gravitationally, the gas perturbations are gradually damped, too.  This fundamental physical difference leads us to suspect that two-fluid prescriptions are not trustworthy in describing coupled star-gas dynamics.

Note that in the top three rows, the strength of swing amplification decreases significantly by the time we get to the rightmost panel ($r=c/\sigma =0.4$).
This is because at a fixed  \textit{fixed} $Q_\mathrm{s}$ and $Q_\mathrm{g}$, a reduction in, e.g., the sound speed $c$ also means a reduction in $\Sigma_\mathrm{g0}$, to which the linear response is always proportional. Another way to understand this is look at the red symbols in Fig.~\ref{fig:rafikov_dispersion}; as we reduce $r$, the stability boundary moves further away.
As we will see in \S\ref{sec:max_amplification} it is the distance to this boundary that mostly determines the amplification factor.
However, the trend is not necessarily monotonic as we move from left to right in Fig.~\ref{fig:single_k_swing_amplification}, because there are also complicated dependencies on the variables $k_{y_\mathrm{c}}$ and $t_\mathrm{i}$. For instance, the stars in panel (b) respond slightly more vigorously than those in panel (a), and the peak response in panel (l) is about as strong as that in panel (k).
\subsection{Maximum amplification}
\label{sec:max_amplification}

\begin{figure}
    \centering
    \includegraphics[width=0.85\linewidth]{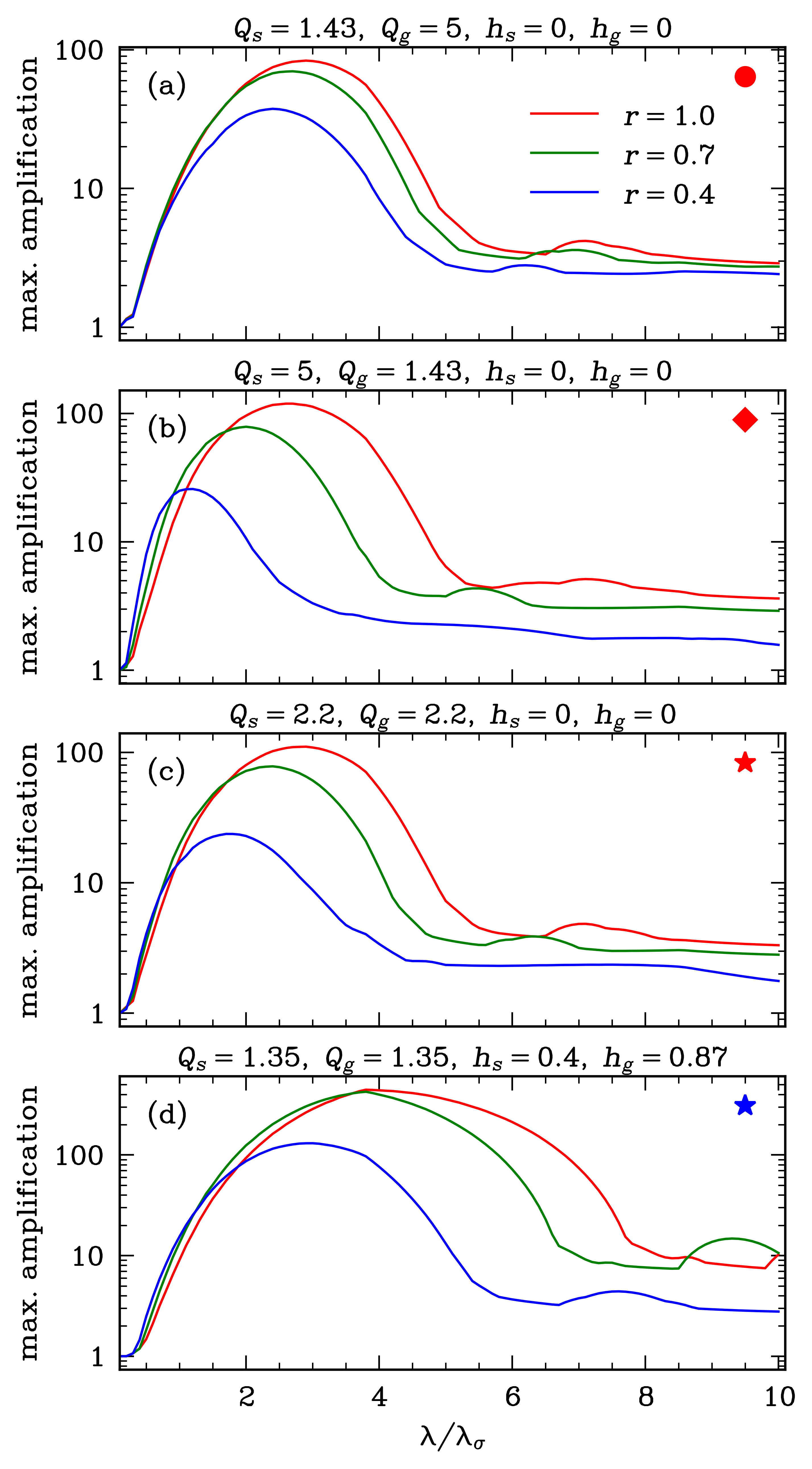}
    \caption{Maximum amplification factor (defined in Eq. \eqref{eqn:max_amp}) of the stellar surface density as a function of dimensionless wavelength $\lambda/\lambda_{\sigma}$, for the same values of $\Qs,\Qg,r,\hs,\hg$ used in Fig.~\ref{fig:single_k_swing_amplification}.
    }
    \label{fig:maximum_amplification}
\end{figure}
\begin{figure*}[thp!]
    \centering
    \includegraphics[width=0.4\linewidth]{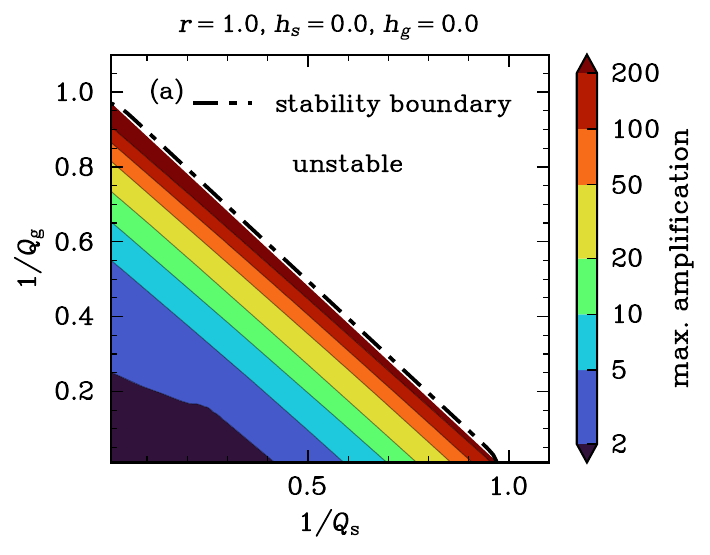}
    \includegraphics[width=0.4\linewidth]{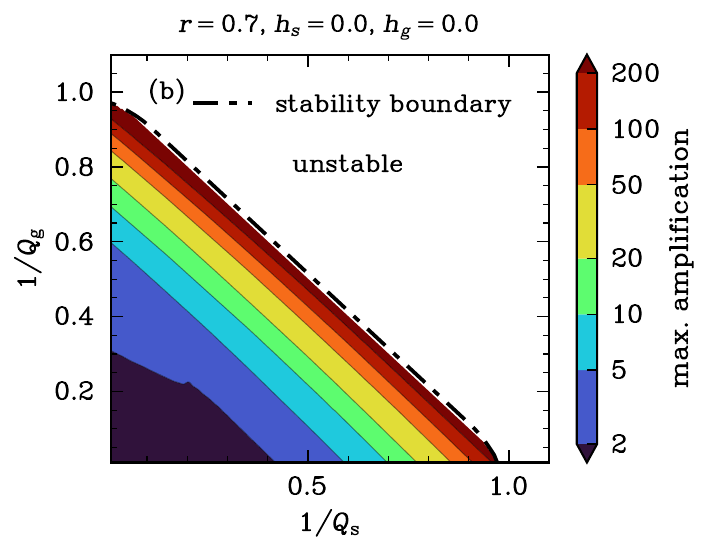}
    \includegraphics[width=0.4\linewidth]{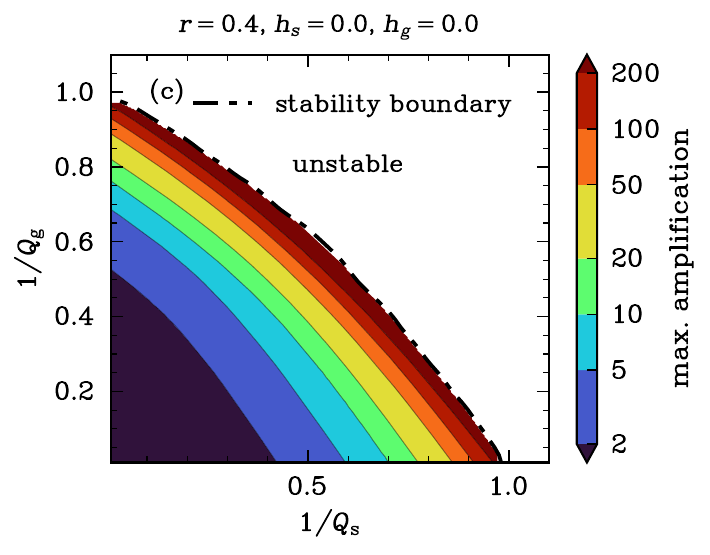}
    \includegraphics[width=0.4\linewidth]{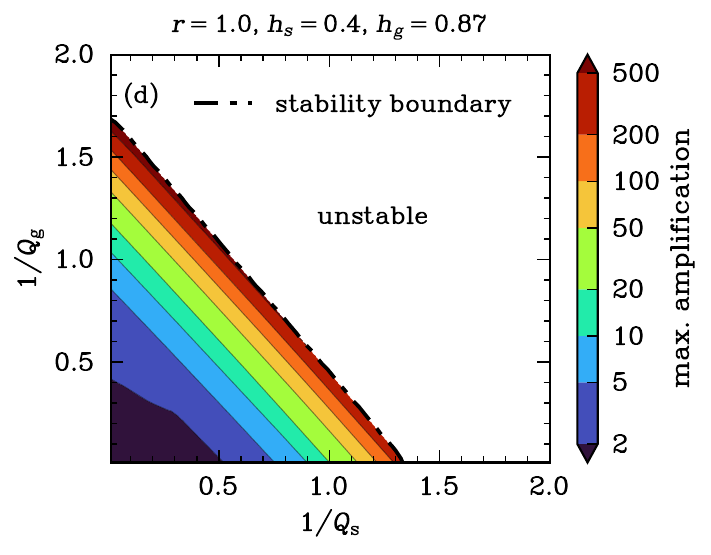}
    \caption{Maximum amplification achievable in the stable regions of the $(1/Q_{\rm s}, 1/Q_{\rm g})$ plane for fixed values of $r, \hs, \hg$, calculated by maximizing the ratio \eqref{eqn:max_amp} over $\lambda/\lambda_{\sigma}$, or equivalently  extracting the peak value from curves like those in Fig.~\ref{fig:maximum_amplification}. The dashed black line denotes the local axisymmetric stability boundary according to the criterion \eqref{eq:rafikov-stability}.}
    \label{fig:maximum_amplification_grid}
\end{figure*}

We now quantify how large a response can be obtained when the initial time $t_\mathrm{i}$ and/or azimuthal wavelength $\lambda = 2\pi/k_{y_\mathrm{c}}$ are allowed to vary. To do this, 
we excite the disk with the same impulsive surface density perturbation used in \S\ref{sec:single}.  We then compute the largest absolute stellar surface density response produced by the full self-gravitating star--gas system, and normalize it by the largest response of a passive, non-self-gravitating stellar sheet to the same impulse:
\begin{align}
    \mathrm{max. \,\,amplification}
    \equiv
    \frac{
    \displaystyle
    \max_{t_\mathrm{i}}
    \max_{t> t_\mathrm{i}}
    \left|
    \delta \hat{\Sigma}_{\mathrm{s}}
    (t; t_\mathrm{i})
    \right|
    }{
    \displaystyle
    \max_{t_\mathrm{i}}
    \max_{t> t_\mathrm{i}}
    \left|
    \delta \hat{\Sigma}_{\mathrm{s}}^{0}
    (t; t_\mathrm{i})
    \right|
    }.
    \label{eqn:max_amp}
\end{align}
Here $\delta \hat{\Sigma}_{\mathrm{s}}^{0}$ denotes the passive response of the stellar component to the external perturbation, without allowing gravitational coupling either to itself or to the gas.
With this normalization, a maximum amplification factor of $\sim 1$ means that self-gravity has not significantly enhanced the stellar response, while a factor $\gg 1$ indicates strong swing amplification.

In Fig.~\ref{fig:maximum_amplification} we plot the maximum amplification factor \eqref{eqn:max_amp} as a function of normalized azimuthal wavelength $\lambda/\lambda_{\sigma}$, where $\lambda_{\sigma} \equiv 2\pi/k_{\sigma}$, for the same twelve sets of the dimensionless parameters  $(\Qs,\Qg,r,\hs,\hg)$ used to create Fig.~\ref{fig:single_k_swing_amplification}. 
For the three thin disk cases (panels (a)-(c)) the amplification curves are rather similar in shape. In particular, in each of these panels, as $r$ is decreased, the maximum amplification decreases.
This is expected since the stability boundary moves further away as $r$ is decreased (see Fig.~\ref{fig:rafikov_dispersion}). 
The maximally-amplified wavelength $\lambda$ also decreases relative to $\lambda_\sigma$. For the thicker disk (panel (d)) the amplification curves show the same qualitative trends but are of quantitatively different shapes. 
In particular, the smallest-wavelength response is suppressed as expected from Eq. \eqref{eqn:thick-midplane-poisson}.

Finally we calculated the peak value of these amplification curves --- i.e., we maximized the dimensionless ratio on the right hand side of \eqref{eqn:max_amp} over the values of $\lambda/\lambda_{\sigma}$ --- 
for every stable location (according to Eq. \eqref{eq:rafikov-stability}) in the $(1/\Qs,1/\Qg)$ plane for various fixed values of $r, \hs, \hg$.   The resulting contour plots are shown in Fig.~\ref{fig:maximum_amplification_grid}.
In each case the stability boundary \eqref{eq:rafikov-stability} is shown with a black dot-dashed curve. 
In all four panels, we see that contours of maximum (non-axisymmetric) swing amplification lie almost parallel to the (axisymmetric) stability curves. 
Thus, the distance from the stability boundary \textit{in the $(1/\Qs,1/\Qg)$ plane} is what primarily determines the strength of swing amplification. 
Within a strip of width $\sim 0.1$ on the stable side of the boundary, amplification factors of $\gtrsim 100$ are commonplace, while disks far from the boundary exhibit much weaker responses to a given perturbation.

\section{Discussion and conclusion}
\label{sec:discussion}

In this paper we have studied  the linear theory of swing amplification of a shearing sheet consisting of both a (collisionless) stellar component and a gaseous component. 
We wrote down the coupled linear perturbation equations for the combined star-gas disk in the case of both 2D (\S\ref{sec:2D}) and (effective) 3D (\S\ref{sec:3D}) disks. 
In the axisymmetric limit (\S\ref{sec:axisymmetric}) we recovered the standard local stability criterion (Eq. \eqref{eq:rafikov-stability}).
We then studied impulsively excited, non-axisymmetric swinging waves in stable disks (\S\ref{sec:single_wave}). 
 
The first key result of this paper is not surprising: strong non-axisymmetric structure can arise in linearly stable star--gas disks through transient swing amplification, provided the disk lies close to the axisymmetric stability boundary in the $(1/\Qs,1/\Qg)$ plane (Fig.~\ref{fig:maximum_amplification_grid}). 
In that sense, the Rafikov/Kim--Ostriker axisymmetric criterion \eqref{eq:rafikov-stability} is useful not only as a stability test, but also as a fair guide to when one should expect spirals and bars to arise in high redshift disks with large gas fractions \citep{blandhawthorn2024,george2025redefining,ghosh2025spiral}. 
Finite-thickness disks behave qualitatively similarly to razor-thin ones, but quantitatively the stability boundary shifts and finite disk thickness suppresses amplification on small scales (Fig.~\ref{fig:maximum_amplification}). 

Our second key result is that the swing amplification in real star-gas disks differs qualitatively from that in `effective two-fluid' disks where stars are treated as another gaseous component (e.g., \citealt{jog1992swing,ghosh2018effect}). 
That is because in two-fluid disks without viscosity, there is no way for energy to be dissipated, meaning such systems can support non-axisymmetric swinging waves whose amplitude is purely oscillatory. In our framework, phase mixing of the stars' distribution function acts as a sink of perturbation energy, so individual swinging waves amplify transiently and then decay: there are no neutral non-axisymmetric modes. Solving the true coupled star-gas equations  \eqref{eq:master-eqn-matrix-form} is not significantly more computationally expensive than solving the more approximate two-fluid equations, so we believe the former approach is always preferable. 

A recent study that is closely related to our work is the paper by \cite{george2025redefining}. These authors investigated whether one could define an `effective $Q$' parameter that combines information about both the axisymmetric stability criterion (\S\ref{sec:axisymmetric}) and the maximum amplification factor (\S\ref{sec:max_amplification}) into a single number for a multicomponent disk.  Both our works share the same conviction that the `distance to the axisymmetric stability boundary' is what determines the non-axisymmetric maximum amplification factor. 
Still, \cite{george2025redefining}'s approach has some advantages and some disadvantages compared to ours. On the one hand, their simple analytic formulae can be rapidly evaluated for disks with arbitrarily many stellar and gaseous components. By contrast we limited ourselves to just two components, and while our approach could easily be generalized, the computational cost of solving the generalized Eq. \eqref{eq:master-eqn-matrix-form} would soon become prohibitive. 
On the other hand, in deriving their results for systems with more than one component, \cite{george2025redefining} treated the stars with a simplified fluid-like prescription following \cite{toomre1981amplifies}\footnote{Note, though, that this is not the same as the effective fluid model used by \cite{jog1992swing}. Instead, \cite{toomre1981amplifies} derives the linear perturbation equations for stars on initially circular orbits, and then includes the effect of finite velocity dispersion at the end in an ad-hoc manner by multiplying by the LSK reduction factor.}, rather than using the linearized collisionless Boltzmann equation as we have done here.  Thus, the intricate physics of swing amplification with a true collisionless component is likely missing from their models. 
In cases with only one stellar and one gaseous component we would advocate solving the true star-gas equations directly, and, if desirable, reading off the maximum amplification factor and most-amplified wavelength from plots like Figs. \ref{fig:maximum_amplification}-\ref{fig:maximum_amplification_grid}. These are easy to generate with our \texttt{python} code \citep{github_code}.

Finally, while our primary motivation stems from recent JWST and ALMA observations of non-axisymmetric structures at $z \simeq 4$, there is an undeniable physical gap between the turbulent, multiphase, feedback-driven chaotic reality of these early galaxies and our idealized study. We relied here on linear perturbations to a smooth background with a constant effective sound speed and treated finite thickness only approximately. We are completely ignorant of nonlinear phenomena \citep{Kim2007} that might occur once $\vert \delta\Sigma_{i}/\Sigma_{i0} \vert \sim 1$ (e.g., shocks, trapping, mode coupling). Furthermore, the local approximation means our approach cannot capture intrinsically global effects such as mode reflection between resonances, bar formation, etc. We also neglected magnetic fields; these will act on gas but not on stars, and so may destroy the phase relationships that here led to gas and stars cooperating gravitationally. 
Some or all of these simplifications may need to be removed before we can make confident astrophysical predictions about combined star-gas dynamics. 
Nevertheless, we now have a consistent (semi-)analytic baseline upon which a better theory can be built.


\begin{acknowledgements}
We thank Eliot Quataert, Scott Tremaine and Joss Bland-Hawthorn for useful conversations.  C.H. was partly supported by the John N. Bahcall Fellowship Fund at the Institute for Advanced Study.
\end{acknowledgements}


\begingroup
  \let\newpage\relax
  \let\clearpage\relax
  \bibliographystyle{aasjournal}
  \bibliography{ref}{}
\endgroup

\appendix
\section{Derivation of master equations}
\label{sec:appendix_derivation}

When we ignore terms of second and higher order in the small parameters $\vert x\vert /\overline{R}$,  $\vert y\vert/\overline{R}\ll 1$, 
the continuity and momentum equations for the gas are~\citep{Ryu-Goodman-1992}:
\begin{align}
\label{eq:fluid-eom-1}
    &\partial_t \Sg + \partial_x \left( \Sg v^x \right) +  \partial_y \left( \Sg v^y \right)  = 0\,,\\
    &D v^x - 2 v^y \Omega + \frac{\partial_x \Pg}{\Sg} - 4 \Omega A x = -\partial_x \Phi_{\mathrm{tot}} \,,\\
\label{eq:fluid-eom-3}
    &D v^y + 2 v^x \Omega + \frac{\partial_y \Pg}{\Sg} = -\partial_y \Phi_{\mathrm{tot}} \,,
\end{align}
where $D \equiv \partial_t + v^x \partial_x + v^y \partial_y$ is the material derivative, $\Sg$ is the gas surface density, and $P_\mathrm{g}$ is the gas pressure.
Meanwhile, to the same level of approximation the motion of an individual star is governed by the Hamiltonian~\citep{Binney_2020}:
\begin{align}\label{eq:Hamiltonian-Local}
    H = \frac{\Delta _y^2}{2}-2 \Delta _y \Omega x+\frac{p_x^2}{2}
    +
    \frac{\kappa ^2 x^2}{2}
    +
    \Phi_{\mathrm{tot}}(x,y)
    \,,
\end{align}
where $p_x \equiv v^x$ and $p_y = \overline{R}\Omega + 2\Omega x + v^y$ are the canonical momenta, and
\begin{align}
    \Delta_{y} \equiv p_{y} - \overline{R} \Omega = v^y + 2 \Omega x.
\end{align}
The stellar distribution function $f(x,y,p_x,p_y,t)$ then satisfies the collisionless Boltzmann equation $\partial f/\partial t + [f,H]=0$ where $[\cdot,\cdot]$ is a Poisson bracket. Explicitly, 
\begin{align}\label{eq:Vlasov-shearing-sheet}
    &\partial_t f
    +
    \frac{\partial f}{\partial x}
    p_x
    +
    \frac{\partial f}{\partial y}
    \left[\Delta_y - 2 \Omega x\right]
    \nonumber\\
    &+
    \frac{\partial f}{\partial p_x}
    \bigg[ 2 \Delta _y \Omega - \kappa ^2 x - \partial_x \Phi_{\mathrm{tot}}  \bigg]
    +
    \frac{\partial f}{\partial p_y}
    \bigg[-\partial_y \Phi_{\mathrm{tot}} \bigg]
    =
    0\,.
\end{align}

We now linearize equations \eqref{eq:fluid-eom-1}-\eqref{eq:fluid-eom-3},  \eqref{eq:Vlasov-shearing-sheet} and \eqref{eq:Poisson-eqn} around the background state described in \S\ref{sec:2D}. The resulting equations are
\begin{align}
    &\partial_t \delta \Sg - 2 A x \partial_y \delta \Sg + \Sigma_{\mathrm{g}0} \left( \partial_x u + \partial_y v  \right) = 0\,,\\
    &\partial_t u - 2 A x \partial_y u - 2 \Omega v + \frac{1}{\Sigma_{\mathrm{g}0}}\partial_x \delta \Pg = -\partial_x \delta \Phi_{\mathrm{tot}} \,,\\
    &\partial_t v - 2 B  u - 2 A x \partial_y v + \frac{1}{\Sigma_{\mathrm{g}0}}\partial_y \delta \Pg  = -\partial_y \delta \Phi_{\mathrm{tot}}  \,,\\
    &\partial_t \delta f
    +
    \frac{\partial \delta f }{\partial x}
    p_x
    +
    \frac{\partial \delta f}{\partial y}
    \left[\Delta_y - 2 \Omega x\right]
    \nonumber 
    \\&\quad\quad+
    \frac{\partial \delta f}{\partial p_x}
    \bigg[ 2 \Delta _y \Omega - \kappa ^2 x \bigg]
    \nonumber\\
    &\quad \quad-
    \frac{\partial f_0}{\partial p_x}
    \partial_x \delta \Phitot
    -
    \frac{\partial f_0}{\partial p_y}
    \partial_y \delta \Phitot
    =
    0\,, \label{eqn:lin_Vlas}\\
    &\partial_x^2 \delta \Phi + \partial_y^2 \delta \Phi + 
    \partial_z^2 \delta \Phi = 4 \pi G \left(\delta \Sg + \delta \Ss \right)  \delta(z)\,,
\end{align}
where $\delta $ denotes the perturbation and 
\begin{align}
    u \equiv \delta v^x \,, v \equiv  \delta v^y\,.
\end{align}
We further assume that the pressure and density perturbations are related via Eq. \eqref{eqn:isothermal}.

Next we perform the transformation to shearing coordinates [Eq.~\eqref{eq:comoving-frame-transform}].
We also perform one additional coordinate transform on the linearized collisionless Boltzmann equation \eqref{eqn:lin_Vlas}, namely we eliminate $p_y$ in favor of 
\begin{align}
    w \equiv \Delta_y + 2 B x = p_y - \overline{R} \Omega + 2 B x = v^y-v^y_0.
\end{align}
Then we have
\begin{align}
    &\partial_t \delta \Sg 
    +
    \Sgo
    \left[ 
    \left(\partial_{x_{\mathrm{c}}} + 2 A t \partial_{y_\mathrm{c}} \right)u
    +
    \partial_{y_\mathrm{c}} v
    \right]
    = 0\,,\\
    &\partial_t u - 2 \Omega v = 
    -\left[\partial_{x_{\mathrm{c}}} + 2 A t \partial_{y_\mathrm{c}} \right] \delta \Phi_{\mathrm{tot}} 
    \nonumber\\
    &\quad\quad- c^2 \left[\partial_{x_{\mathrm{c}}} + 2 A t \partial_{y_\mathrm{c}} \right] \frac{ \delta \Sg}{\Sgo}
    \,,\\
    &\partial_t v - 2 B  u = -\partial_{y_\mathrm{c}} \delta \Phi_{\mathrm{tot}} - \frac{c^2 \partial_{y_\mathrm{c}} \delta \Sg}{\Sgo}  \,,\\
    \label{eq:Vlasov-comoving}
    &\partial_t \delta f
    +
    \frac{\partial \delta f }{\partial x_{\mathrm{c}}}
    p_x
    +
    \frac{\partial \delta f}{\partial y_\mathrm{c}}
    \left[w + 2 A t p_x\right]
    +
    2 B p_x
    \frac{\partial \delta f }{\partial w}
    \nonumber\\
    &\quad\quad+
    2 \Omega w
    \frac{\partial \delta f}{\partial p_x} 
    -
    \frac{\partial f_0}{\partial p_x}
    \left(\partial_x + 2 A t \partial_{y_\mathrm{c}} \right)\delta \Phi_{\mathrm{tot}}
    \nonumber\\
    &\quad\quad-
    \frac{\partial f_0}{\partial w}
    \partial_{y_\mathrm{c}} \delta \Phi_{\mathrm{tot}}
    =
    0
    \,, \\
    &\left[\partial_{x_{\mathrm{c}}} + 2 A t \partial_{y_\mathrm{c}} \right]^2 \delta \Phi + \partial_{y_\mathrm{c}}^2 \delta \Phi + 
    \partial_z^2 \delta \Phi \nonumber\\
    &
    \hspace{2cm} 
    = 4 \pi G \left(\delta \Sg + \delta \Ss \right)  \delta(z)\,.
\end{align}
We now Fourier transform [see Eq.~\eqref{eq:fourier-transformation-convention}], with the result
\begin{align}
    \label{eq:cont}
    &\partial_t \dhSg 
    +
    i
    \Sgo
    \left[ 
    \kzero(t) \hu
    + 
    k_{y_\mathrm{c}} \hv
    \right]
    = 0\,,\\
    \label{eq:momen-u}
    &\partial_t \hu - 2 \Omega \hv = 
    -i \kzero(t) 
    \left[\delta \hat{\Phi}_{\mathrm{tot}}
    + c^2\frac{\dhSg }{\Sgo}
    \right]
    \,,\\
    \label{eq:momen-v}
    &\partial_t \hv - 2 B \hu = -i k_{y_\mathrm{c}}
    \left[\delta \hat{\Phi}_{\mathrm{tot}} +\frac{c^2 \dhSg }{\Sgo} \right]
    \,,\\
\label{eq:Vlasov-comoving-FT}
    &\partial_t \dhf
    +
    i \dhf
    \left[
    p_x \kzero(t)
    + k_{y_\mathrm{c}}
    w 
    \right]
    +
    2 \Omega w
    \frac{\partial \dhf }{\partial p_x}
    +
    2 B p_x
    \frac{\partial \dhf }{\partial w}
    \nonumber\\
    &\quad \quad =
    i \kzero(t)
    \frac{\partial f_0}{\partial p_x}
    \delta \hat{\Phi}_{\mathrm{tot}}
    +
    i k_{y_\mathrm{c}}
    \frac{\partial f_0}{\partial w}
    \delta \hat{\Phi}_{\mathrm{tot}}
    \,, \\
    &- \left(\kzero(t)^2 + k_y^2\right) \delta \hat{\Phi}
    +
    \partial_z^2 \delta \hat{\Phi}
    =
    4 \pi G \left( \dhSg + \dhSs \right)  \delta(z)
    \,,
    \label{eqn:Poisson_Fourier}
\end{align}
where 
\begin{equation}
    k_0(t) = k_{x_{\mathrm{c}}}  + 2 A k_{y_{\mathrm{c}}} t.
\end{equation}
When we integrate the first four of these equations over $z$ they are unchanged. Integrating Eq.~\eqref{eqn:Poisson_Fourier} with vertical decay boundary conditions and applying the jump condition across $z=0$ gives Eq.~\eqref{eqn:poisson-solution}, with $k(t)$ defined in \eqref{eqn:koft}.

\subsection{Master equation for gas}

Let us now obtain the master equation  for the gas sector.
First,  Eqs.~\eqref{eq:momen-u} and \eqref{eq:momen-v} imply that
\begin{align}
    \partial_t
    \left(k_{y_\mathrm{c}} \hu - \kzero(t) \hv\right)
    +
    2 B
    \left(k_{y_\mathrm{c}} \hv + \kzero(t) \hu \right)
    =0\,.
\end{align}
Substituting this in the continuity equation [Eq.~\eqref{eq:cont}], we get
\begin{align}
    \partial_t
    \left[ 
    \dhSg
    -
    \frac{i \Sgo}{2B}
    \left(
    k_{y_\mathrm{c}} \hu - \kzero(t) \hv
    \right)
    \right]
    =
    0\,.
\end{align}
This equation represents a statement of the conservation of vorticity.  
Since we always start with unperturbed disks in this paper, the constant of integration is zero.
Using the resulting identity and eliminating $u$, $v$ and $\dhSg$ using Eqs.~\eqref{eq:cont} and \eqref{eqn:poisson-solution}, we obtain a harmonic oscillator-like equation for the gravitational potential of the gas, Eq. \eqref{eq:gas-master-eqn}, with spring constant
\begin{align}
    S^2(t) 
    &\equiv 
     \kappa^2 - 2 \pi G \Sigma_\mathrm{g0} k(t) + c^2 k^2(t) 
     \nonumber
     \\
     &\quad\quad\quad\quad\quad+ \frac{12 A^2 k_{y_\mathrm{c}}^4} {k^4(t)}
    - \frac{8 A \Omega k_{y_\mathrm{c}}^2}{k^2(t)}.
    \label{eqn:spring_constant}
\end{align}

Note that there is a mistake in the oscillator equation for gas perturbations reported by \cite{GT-78}. Namely, the right hand side of their Eq. (35a) should be multiplied by a factor $-(1+\tau^2)$. However, their Eq. (34) is correct and is equivalent to our Eq. \eqref{eq:gas-master-eqn} if we set $\delta \hat{\Phi}_{\mathrm{s}}=0$. Since they employ Eq. (34) for the remainder of their paper, the mistake does not affect any of their subsequent results.

\subsection{Master equation for stars}

Finally, we obtain an analogous master equation for the stars. Introducing
\begin{align}
    U \equiv \frac{i w \kzero(t)}{2B}
    -
    \frac{i k_{y_\mathrm{c}} p_x}{2 B}
\end{align}
we can simplify Eq.~\eqref{eq:Vlasov-comoving-FT}
to 
\begin{align}
    &\partial_t \left(e^U \dhf\right)
    +
    2 \Omega w
    \frac{\partial  }{\partial p_x}
    \left(e^U \dhf\right)
    +
    2 B p_x
    \frac{\partial }{\partial w}
    \left(e^U \dhf\right)
    \nonumber\\
    &=
    e^U
    \bigg[
    i \kzero(t)
    \frac{\partial f_0}{\partial p_x}
    \delta \hat{\Phi}_{\mathrm{tot}}
    +
    i k_{y_\mathrm{c}}
    \frac{\partial f_0}{\partial w}
    \delta \hat{\Phi}_{\mathrm{tot}}
    \bigg]
    \,, \\
    &=
    - e^{U}
    \frac{i f_0}{\sigma^2}\delta \hat{\Phi}_{\mathrm{tot}}
    \left[p_x \kzero(t) +  \frac{4 k_{y_\mathrm{c}} \Omega^2 w }{\kappa^2}\right]
    \,,
\end{align}
where to get the last line we plugged in the explicit unperturbed distribution function Eq.~\eqref{eq:Vlasov-equilibrium}.
The characteristics of this equation are
\begin{align}
    &p_{x,\mathrm{char}}(s)
    =
    p_x \cos [\kappa  (s-t )] + \frac{2 \Omega w \sin [\kappa  (s-t )]}{\kappa }
    \,, \\
    &w_{\mathrm{char}}(s)
    =
    w \cos [\kappa  (s-t )]
    -
    \frac{\kappa  p_x \sin [\kappa  (s-t )]}{2 \Omega}
    \,.
\end{align}
so the solution is
\begin{align}
    &\delta \hat{f}(t,w,p_x)
    =
    - 
    e^{-U(t)}
    \int_{t_\mathrm{i}}^{t} ds\,
    e^{U[w_\mathrm{char}(s), p_{x,\mathrm{char}}(s), s]} \nonumber \\ &\times 
    \frac{i f_0}{\sigma^2}\delta \hat{\Phi}_{\mathrm{tot}}(s)
    \left[p_{x,\mathrm{char}}(s) \kzero(s) +  \frac{4 k_{y_\mathrm{c}} \Omega^2 w_{\mathrm{char}}(s) }{\kappa^2}\right]
    \,,
\end{align}
where we have assumed that $\delta \hat{f}(t=t_\mathrm{i},w,p_x) = 0$.
Using this equation, we can calculate the perturbed stellar surface density
\begin{align}
    \dhSs
    =
    \int_{-\infty}^\infty dp_x \int_{-\infty}^\infty d w \,
    \delta \hat{f}(t,w,p_x)
    \,.
\end{align}
The end result is
\begin{align}
    \dhSs(t)
    =
    \int_{t_\mathrm{i}}^{t} ds \kappa \, K(t,s) \left[\dhSs(s) + \dhSg(s) + \delta \hat{\Sigma}_{\mathrm{ext}}(s) \right],
\end{align}
where we have defined the JT kernel
\begin{align}\label{eq:JT-Kernel}
    K(t,s) \equiv \frac{8 \pi G \Sso}{\kappa}
    \frac{k_{y_\mathrm{c}}}{k(s)} r_1 e^{{-{\sigma^2 r_2}/{\kappa^4}} },
\end{align}
where 
\begin{align}
    r_1 &\equiv \frac{Y_1 \cos[\kappa  (s-t)]}{8 B}+\frac{Y_2 \sin[\kappa  (s-t)]}{16 B \kappa  k_{y_\mathrm{c}}\Omega}
    \,,\\
    r_2 &\equiv 
    -Y_2 \cos[\kappa  (s-t)]+2 \kappa  k_{y_\mathrm{c}}\Omega Y_1 \sin[\kappa  (s-t)]
    \nonumber\\
    &+\frac{\kappa ^2 Y_1^2}{2}+Y_2
    \,,
\end{align}
and
\begin{align}
    Y_1 &\equiv k_0(t) - k_0(s) \,,\\
    Y_2 &\equiv \kappa ^2 k_0(s) k_0(t)+4 k_{y_\mathrm{c}}^2 \Omega^2 \,.
\end{align}
Finally we can use \eqref{eqn:poisson-solution} to convert this from an equation for density fluctuations into an equation for potential fluctuations. The result is Eq.~\eqref{eq:stars-master-eqn} with kernel
\begin{align}
    \Kpot(t,s) \equiv \frac{\kappa K(t,s) k(s)}{k(t)}.
    \label{eqn:potential_kernel}
\end{align}

\subsection{Including finite disk thickness}\label{appendix:thick-disk}

As discussed in \S\ref{sec:3D} it is possible to perform an effective finite-thickness generalization of the shearing sheet equations using the formula \eqref{eqn:thick-midplane-poisson} for midplane potential sourced by species $i$. 
Following closely analogous calculations to those above, one ultimately arrives at the following master equations
\begin{align}
    &\partial_t^2 \delta \hat{\Phi}_{\mathrm{g}} 
    +
    \partial_t \delta \hat{\Phi}_{\mathrm{g}} \frac{S_1(t)}{{1+H_{\mathrm{g}} k(t)}}
    +
    \frac{S^2_{\mathrm{thick}}(t)}{1+H_{\mathrm{g}} k(t)}
    \delta \hat{\Phi}_{\mathrm{g}}
    \nonumber\\
    &=
    \frac{2\pi G \Sgo k(t)}{1 + H_\mathrm{g} k(t)}
    \left[ 
    \delta \hat{\Phi}_{\mathrm{s}}
    +
    \delta \hat{\Phi}_{\mathrm{ext}}
    \right]
    \,,\\
    &\delta \hat{\Phi}_\mathrm{s}(t)
    =
    \int_{t_\mathrm{i}}^t ds \, K_{\mathrm{pot,thick}}(t,s) 
    \bigg[ 
    \delta \hat{\Phi}_{\mathrm{s}}(s)
    \nonumber\\
    &
    +
    \delta \hat{\Phi}_{\mathrm{g}}(s)
    +
    \delta \hat{\Phi}_{\mathrm{ext}}(s)
    \bigg]
    \,,
\end{align}
where
\begin{align}
    &S_1(t)
    =
    \frac{4 A H_{\mathrm{g}} k_{\yc} k_0(t)}{k(t)}
    \,,\\
    &S^2_{\mathrm{thick}}(t)
    =
    S^2(t)
    +
    H_{\mathrm{g}}
    \bigg[
    \frac{16 A^2 k_{\yc}^4}{k(t)^3}-\frac{2 A k_{\yc}^2 \left(4 A \Omega+\kappa ^2\right)}{\Omega k(t)}
    \nonumber\\
    &\quad 
    +c^2 k(t)^3+\kappa ^2 k(t)
    \bigg]
    \,,\\
    &K_{\mathrm{pot,thick}}(t,s) = \frac{K_\mathrm{pot}(t,s)}{1+ H_{\mathrm{s}} k(t)}.
\end{align}
These of course reduce to the 2D equations (\S\ref{sec:2D}) in the limit of $H_\mathrm{s}=H_\mathrm{g}=0$.

\section{Numerical scheme}\label{appendix:num-scheme}

To solve   equations \eqref{eq:gas-master-eqn}-\eqref{eq:stars-master-eqn} numerically, we first express them in a compact integral form as
\begin{align}\label{eq:master-eqn-matrix-form}
    \boldsymbol{\Psi}(t)
    &=
    \boldsymbol{\Psi}(t=t_\mathrm{i})
    +
    \int_{t_\mathrm{i}}^{t}
    dt' \boldsymbol{A}(t,t')\cdot\boldsymbol{\Psi}(t')
    \nonumber\\
    &+
    \int_{t_\mathrm{i}}^{t} dt' \boldsymbol{G}(t,t') \cdot \boldsymbol{F}_{\mathrm{ext}}(t')
    \,,
\end{align}
where
\begin{align}
    &\boldsymbol{\Psi}(t)
    =
    \left[\delta \hat{\Phi}_{\mathrm{g}} (t), \partial_t \delta \hat{\Phi}_{\mathrm{g}}(t), \delta \hat{\Phi}_{\mathrm{s}}(t) \right]^{T}
    \,,\\
    &\boldsymbol{A}(t,t')
    =
    \begin{pmatrix}
        0 & 1 & 0 \\
        -S^2(t') & 0 & 2 \pi G \Sgo k(t') \\
        \Kpot(t,t') & 0 & \Kpot(t,t')
    \end{pmatrix}
    \,,\\
    &\boldsymbol{F}_{\mathrm{ext}}(t') = 
    \left[0, 0, \delta \hat{\Sigma}_{\mathrm{ext}} (t') \right]^T
    \,,\\
    &\boldsymbol{G}(t,t') = 
    \begin{pmatrix}
        0 & 0 & 0 \\
        0 & 0 &  - (2 \pi G)^2 \Sgo  \\
        0 & 0 & - {2 \pi G \Kpot(t,t')}/{k(t')}
    \end{pmatrix}
    \,.
\end{align}
We then discretize this equation on a uniform grid of time values with step-size $h$, and letting
\begin{align}
    t_n \equiv t_\mathrm{i} + nh\,.
\end{align}
The numerical scheme we follow is a simple trapezoidal rule:
\begin{subequations}\label{eq:numerical-scheme}
\begin{align}
    &\boldsymbol{\Psi}(t_{n}) =  \frac{h}{2} \boldsymbol{A}(t_{n}, t_\mathrm{i})\cdot \boldsymbol{\Psi}(t_\mathrm{i})
    +
    h \sum_{k=1}^{n-1} \boldsymbol{A}(t_{n}, t_k)\cdot \boldsymbol{\Psi}(t_k) 
    \nonumber\\
    &+
    \frac{h}{2} \boldsymbol{A}(t_{n}, t_{n})\cdot \boldsymbol{\Psi}(t_{n})
    +
    \boldsymbol{B}(t_{n})
    \,,
\end{align}
\end{subequations}
where
\begin{align}
    \boldsymbol{B}(t) = \int_{t_\mathrm{i}}^{t} dt' \boldsymbol{G}(t,t') \cdot \boldsymbol{F}_{\mathrm{ext}} \,.
\end{align}
Given this discretization, finding the solution for $\boldsymbol{\Psi}$ at every timestep becomes a problem in linear algebra. 

When we solve the effective 3D equations (\S\ref{sec:3D}), or the two-fluid analogue of the 2D equations (\S\ref{sec:single}), we follow a procedure completely analogous to that just described, first putting the equations in integral form and then discretizing them on a temporal grid.
The only difference is the definition of the vectors $\bm{A}$ and $\bm{G}$.

\section{Response to a cloud}
\label{appendix:cloud_details}
In this appendix, we describe how to calculate the time-asymptotic linear response of the star-gas disk to a cloud.
Let us first note that the cloud is stationary in the physical coordinates $(x,y)$. However, it is explicitly time-dependent when written in the shearing coordinates $(x_\mathrm{c},\yc)$ (Eq.~\ref{eq:comoving-frame-transform}). Our goal is to compute the stationary response of the gaseous and stellar surface densities in $(x,y)$.

Using the Fourier convention of Eq.~\eqref{eq:fourier-transformation-convention}, the Fourier transformed  external density is
\begin{align}
    \delta\hat{\Sigma}_{\mathrm{ext}}(k_{x_\mathrm{c}},k_{\yc},t)
    =
    \frac{M}{4\pi^2}e^{-{k^2(t)\Delta^2}/{2}}\,,
\end{align}
where $M$ is the mass of the cloud.

We evolve the time-dependent master equation [Eq.~\eqref{eq:master-eqn-matrix-form}] with this forcing, taking the limit $t_\mathrm{i}=-\infty$ and assuming that the initial condition is zero, 
$\boldsymbol{\Psi}(t_\mathrm{i})=0$. In this limit, the master equation  is invariant under the combined transformation
\begin{align}
    t \to t+a
    \,,\qquad
    k_{x_{\mathrm{c}}} \to k_{x_{\mathrm{c}}}-2Ak_{\yc}a
    \,,\qquad
    k_{\yc}\to k_{\yc}
    \,.
    \label{eq:shearing-shift-symmetry}
\end{align}
This is not an ordinary time-translation symmetry at fixed $k_{x_{\mathrm{c}}}$. Instead, it is the symmetry that leaves the physical radial wavenumber $k_0(t)=k_{x_{\mathrm{c}}}+2Atk_{\yc}$ unchanged. 

This invariance allows us to relate the solution at arbitrary sheared radial label $k_{x_{\mathrm{c}}}$ to the solution with $k_{x_{\mathrm{c}}}=0$. Let $\boldsymbol{\Psi}_{k_{x_{\mathrm{c}}},k_{\yc}}(t)$
denote the retarded solution of the master equation for fixed sheared Fourier labels $(k_{x_{\mathrm{c}}},k_{\yc})$. Define the shifted solution
\begin{align}
    \boldsymbol{\Psi}^{(a)}_{k_{x_{\mathrm{c}}},k_{\yc}}(t)
    \equiv
    \boldsymbol{\Psi}_{k_{x_{\mathrm{c}}}-2Ak_{\yc}a,k_{\yc}}(t+a)
    \,.
\end{align}
By the symmetry in Eq.~\eqref{eq:shearing-shift-symmetry}, and because $t_\mathrm{i}=-\infty$, the shifted solution satisfies the same retarded Volterra equation as $\boldsymbol{\Psi}_{k_{x_{\mathrm{c}}},k_{\yc}}(t)$. Assuming uniqueness of the retarded solution, we therefore have
\begin{align}
    \boldsymbol{\Psi}_{k_{x_{\mathrm{c}}},k_{\yc}}(t)
    =
    \boldsymbol{\Psi}_{k_{x_{\mathrm{c}}}-2Ak_{\yc}a,k_{\yc}}(t+a)
    \,.
\end{align}
Choosing
\begin{align}
    a=\frac{k_{x_{\mathrm{c}}}}{2Ak_{\yc}}
\end{align}
sets the shifted radial label to zero, giving
\begin{align}
    \boldsymbol{\Psi}_{k_{x_{\mathrm{c}}},k_{\yc}}(t)
    =
    \boldsymbol{\Psi}_{0,k_{\yc}}
    \left(
        t+\frac{k_{x_{\mathrm{c}}}}{2Ak_{\yc}}
    \right)
    =
    \boldsymbol{\Psi}_{0,k_{\yc}}
    \left(
        \frac{k_0(t)}{2Ak_{\yc}}
    \right)
    \,.
    \label{eq:solution-relabeling}
\end{align}
Equation~\eqref{eq:solution-relabeling} is the key identity used below. It shows that the apparent three-variable dependence on $(t,k_{x_{\mathrm{c}}},k_{\yc})$ can be reduced to a two-variable dependence on $(k_0(t),k_{\yc})$. Equivalently, for a stationary response, the shearing-wave time can be reinterpreted as a label for the physical radial wavenumber.

Let
\begin{align}
    S_i(k_{\yc},\tau)
    \equiv
    \delta\hat{\Sigma}_{i,0,k_{\yc}}(\tau)
    \,,\qquad
    i\in\{g,s\}
    \,,
\end{align}
be the gas or stellar surface density response obtained by solving the master equations with $k_{x_{\mathrm{c}}}=0$.
In the physical coordinates $(x,y)$, we write the stationary response as 
\begin{align}
    \delta\Sigma_i(x,y)
    =
    \int dk_x\,dk_y\,
    \delta\widetilde{\Sigma}_i(k_x,k_y)
    e^{i(k_xx+k_y y)}
    \,,
    \label{eq:physical-fourier-transform}
\end{align}
where $\delta\widetilde{\Sigma}_i(k_x,k_y)$ is the Fourier amplitude in the physical, unsheared coordinates. On the other hand, in the sheared coordinates $(x_c,\yc)$, the same perturbation can be written as (inverting Eq.~\ref{eq:fourier-transformation-convention}):
\begin{align}
    \delta\Sigma_i(x_c,\yc,t)
    =
    \int dk_{x_{\mathrm{c}}}\,dk_{\yc}\,
    \delta\hat{\Sigma}_{i}(k_{x_{\mathrm{c}}},k_{\yc},t)
    e^{i(k_{x_{\mathrm{c}}} x_c + k_{\yc}\yc)}
    \,.
    \label{eq:sheared-fourier-transform}
\end{align}
Using $\yc=y+2Axt$, and equating Eqs.~\eqref{eq:physical-fourier-transform} and \eqref{eq:sheared-fourier-transform} gives
\begin{align}
    \delta\widetilde{\Sigma}_i(k_x,k_y)
    =
    \delta\hat{\Sigma}_{i}(k_{x}-2Atk_{\yc},k_{\yc},t)
    \,.
    \label{eq:physical-sheared-amplitude-relation}
\end{align}

Now using Eq.~\eqref{eq:solution-relabeling}, the sheared-coordinate solution satisfies
\begin{align}
    \delta\hat{\Sigma}_{i}(k_{x_{\mathrm{c}}},k_{\yc},t)
    &=
    \delta\hat{\Sigma}_{i}\left(0,k_{\yc},
        t+\frac{k_{x_{\mathrm{c}}}}{2Ak_{\yc}}
    \right)
    \nonumber\\
    &=
    S_i
    \left(
        k_{\yc},
        t+\frac{k_{x_{\mathrm{c}}}}{2Ak_{\yc}}
    \right)
    \,.
\end{align}
Substituting $k_{x_{\mathrm{c}}}=k_x-2Atk_{\yc}$ into this expression gives
\begin{align}
    t+\frac{k_{x_{\mathrm{c}}}}{2Ak_{\yc}}
    &=
    t+\frac{k_x-2Atk_{\yc}}{2Ak_{\yc}}
    =
    \frac{k_x}{2Ak_{\yc}}
    \,.
\end{align}
Therefore Eq.~\eqref{eq:physical-sheared-amplitude-relation} becomes
\begin{align}
    \delta\widetilde{\Sigma}_i(k_x,k_y)
    =
    S_i\left(k_y,\frac{k_x}{2Ak_y}\right)
    \,,
    \qquad i\in\{g,s\}
    \,.
    \label{eq:stationary-physical-fourier-amplitude}
\end{align}
This identity is the main practical result. It shows that the stationary
Fourier amplitude in the physical coordinates can be obtained from the
$k_{x_{\mathrm{c}}}=0$ shearing-wave calculation by reinterpreting the shearing-wave time
as the physical radial-wavenumber label,
\begin{align}
    \tau = \frac{k_x}{2Ak_y}.
\end{align}

In the numerical calculation, we evaluate
Eq.~\eqref{eq:stationary-physical-fourier-amplitude} on a discrete Fourier
grid. We use a real-space box with
\begin{align}
    -\frac{L_x}{2} < x < \frac{L_x}{2}
    \,,\qquad
    -\frac{L_y}{2} < y < \frac{L_y}{2}
    \,,
\end{align}
and discretize it with $N_x$ and $N_y$ grid points. The use of a finite Fourier box implies periodic boundary conditions: the reconstructed wake is
periodic with periods $L_x$ and $L_y$. Therefore, in practice, we choose
$L_x$ and $L_y$ larger than the region shown in the figures, so that periodic
copies of the cloud wake do not affect the plotted domain.

The continuous inverse Fourier transform is then approximated by the truncated
sum
\begin{align}
    \delta\Sigma_i(x,y)
    &\simeq
    \frac{2\pi}{L_x}
    \frac{2\pi}{L_y}
    \sum_{m=-N_{k_x}/2}^{N_{k_x}/2-1}
    \sum_{n=-N_{k_y}/2}^{N_{k_y}/2-1}
    \delta\widetilde{\Sigma}_i
    \left(
        \frac{2\pi m}{L_x},
        \frac{2\pi n}{L_y}
    \right)
    \nonumber\\
    &\times e^{\left[
        i\left(
            \frac{2\pi m x}{L_x}
            +
            \frac{2\pi n y}{L_y}
        \right)
    \right]}
    \,.
\end{align}
When the sum is evaluated with a discrete Fourier transform on an
$N_x\times N_y$ real-space grid, we take
$N_{k_x}=N_x$ and $N_{k_y}=N_y$.

\end{document}